\documentstyle[pre,aps,floats,twocolumn,epsf]{revtex}

\begin{document}

\def\R{\mbox{$I\!\!R$}}

\def\f{\mbox{\mbf f}}
\def\g{\mbox{\mbf g}}
\def\h{\mbox{\mbf h}}

\def\x{\mbox{\mbf x}}
\def\y{\mbox{\mbf y}}
\def\z{\mbox{\mbf z}}

\def\zero{\mbox{\mbf 0}}

\def\H{\mbox{\mbf H}}

\def\bfphi{\mbox{\mbf \phi}}

\newcommand{\mbf}[1]{\mbox{\boldmath $#1$}}

\newtheorem{definition}{Definition}

\title{A unifying definition of synchronization for dynamical systems} 

\author{Reggie Brown}
\address{Department of Physics and Department of Applied Science, College
of William and Mary, Williamsburg, VA 23187-8795}

\author{Ljup\v{c}o Kocarev}
\address{Department of Electrical Engineering, St Cyril and Methodius 
 University, Skopje, PO Box 574, Macedonia}

\date{\today}
\maketitle

\begin{abstract}
We propose a unifying definition for synchronization.  By example,
we show that the synchronization phenomena discussed in the
dynamical systems literature fits within the framework of this
definition.

\end{abstract}


Synchronization between dynamical systems has been an active research 
topic since the time of 
Huygens.  It is a phenomenon of interest to fields ranging from 
celestial mechanics to laser physics, and from communication to 
neuroscience~\cite{chaos}.  

Over the last decade, a number of new types of synchronization have
appeared: chaotic synchronization~\cite{pc}, phase 
synchronization~\cite{rpk}, lag synchronization~\cite{rpk-1},
and generalized synchronization~\cite{rsta}, to mention only a few.  
This is in addition to the classic examples of synchronization
in periodic systems~\cite{blek,arnold}.  Many of these have been
experimentally observed in a single system~\cite{tdhh}. 
Synchronization is often categorized on the basis of whether the
coupling mechanism is uni-directional or bi-directional.  Stable 
synchronization with uni-directional coupling has been called
synchronization by an external force (for frequency synchronization)
and master-slave synchronization (Pecora and Carroll~\cite{pc}).
(It has recently been shown that, if the synchronized systems
are identical then there is no essential difference between 
uni-directional and bi-directional synchronization~\cite{josic}.)

Although there have been several attempts~\cite{cc,kp}, no
successful definition of synchronization currently exists.  The 
definition in use is an ever increasing enumerated list.  When
a ``new type'' of synchronization arises, its name is added to
the list.  We believe that  ``definition by example'' is an 
untidy situation which should be replaced by a single definition
that encompasses all of the known examples.

In this letter we propose a unified definition which accounts for
all types of synchronization between finite dimensional systems.
Although we explicitly discusses synchronization between two
continuous time dynamical systems, our results can be extended
to $N$ continuous time, or $N$ discrete time systems.  Therefore, 
our results apply to a larger class of phenomena than the one we
explicitly discussed.

To construct this definition, assume that a large stationary
deterministic dynamical system is divided into two sub-systems,
\begin{eqnarray}
\label{eq-system} 
\frac{d \x}{dt} & = & \f_1(\x, \y; t) \\
\frac{d \y}{dt} & = & \f_2(\y, \x; t). \nonumber
\end{eqnarray}
Here, $\x \in \R^{d_1}$ and $\y \in \R^{d_2}$ are vectors that may have
different dimensions.  The phase space and vector field of the large
system is formed (in a natural way) from the product of the two smaller
phase spaces and vector fields.  Examples of phenomena that can be 
described by Eq.~(\ref{eq-system}) are ubiquitous.

Colloquially, synchronization means correlated in-time behavior
between different processes.  In fact, the Oxford Advanced
dictionary~\cite{oxford}, defines synchronization as ``to agree in time''
and ``to happen at the same time''.  From this intuitive definition we
propose that synchronization requires: (1) {\em Separating} the dynamics
of a large dynamical system into the dynamics of sub-systems.
(2) A method for {\em measuring properties} of the sub-systems.
(3) A method for {\em comparing} the properties of the sub-systems.
(4) A criteria for determining whether the properties {\em agree in 
time}.  If they agree then the systems are synchronized.  The
remainder of this letter formalizes this intuitive definition of
synchronization by explicitly addressing each requirement, and applying
the proposed definition to examples.

We begin by separating the dynamics of the large systems into the
dynamics of sub-systems.  Let $\bfphi(\z_0)$ denote a trajectory of 
the large dynamical system, given by Eq.~(\ref{eq-system}), with
initial condition, $\z_0 = [\x_0, \y_0] \in \R^{d_1} \otimes 
\R^{d_2}$.  Respectively, curves $\bfphi_x(\z_0)$ and $\bfphi_y
(\z_0)$ are obtained from this trajectory by projecting away the 
\y\ and \x~components.  We say that $\bfphi_x(\z_0)$ and $\bfphi_y
(\z_0)$ are ``trajectories'' of the first and second sub-systems of
Eq.~(\ref{eq-system}).  In this context we have {\em separated}
the trajectories $\bfphi_x(\z_0)$ and $\bfphi_y(\z_0)$ from $\bfphi
(\z_0)$, rather than constructing $\bfphi(\z_0)$ from $\bfphi_x(\z_0)$
and $\bfphi_y(\z_0)$.

To discuss {\em measuring properties} of the sub-systems let ${\cal
X}$ denote the space of all trajectories of the first sub-system,
and consider a mapping $\g_x: {\cal X} \otimes \R \to \R^k$.  The
first \R\ represents time, and is included so that $\g_x$ can make
explicit reference to time.  We say that the mapping, $\g_x$, is a
{\em property} of the first sub-system.  The image of $[\bfphi_x
(\z_0), t] \in {\cal X} \otimes \R$ under the mapping $\g_x$ is the
result of {\em measuring the property} of the first sub-system.  It
will be denoted by $\g(\x) \in \R^k$.  Similar definitions can be 
made for the second sub-system.  The following examples make these
notions less abstract.

For synchronization, a property of a sub-system that is often of
interest is frequency.  Measuring the property means calculating 
a numerical value for the frequency. Hence, $\omega_x = \g(\x)$.
Other properties of interest are the coordinates of a sub-system
at time $t$.  Measuring the properties means determining numerical
values for the coordinates.  Hence, $\x(t) = \g(\x)$.  
Experimentally, $\g_x$ is the quantity being measured, and $\g(\x)$
is the value of the measurement.  These examples show that a property
can be a long time average, or a quantity whose value depends 
implicitly on time.  Furthermore, the dimension of the measurement,
$k$, can take on different values depending on the property being
measured.  Notice that it is reasonable to say $\g_x$ is a 
property of the first sub-system because $\g(\x)$ is obtained 
without explicitly referring to any other sub-system.

Finally, we discuss the notions of {\em comparing} the properties,
and determining when they {\em agree in time}.  We say the function
$\h: \R^k \otimes \R^k \rightarrow \R^k$ {\em compares} the measured 
properties of the two sub-systems, and the two measurements {\em
agree in time} if and only if $\h[\g(\x), \g(\y)] = \zero$. Below,
a norm is used to determine this last requirement.

With these preliminaries in place, we offer the following
definition for synchronization: \\

\noindent {\bf Definition} {\em The sub-systems in 
Eq.~(\ref{eq-system}) are synchronized on the trajectory $\bfphi
(\z_0)$, with respect to the properties, $\g_x$ and $\g_y$, if 
there is a time independent function $\h: \R^k \otimes \R^k \to
\R^k$ such that}
\begin{displaymath}
\| \h[\g(\x), \g(\y)] \| = 0, 
\end{displaymath} 
where $\| \bullet \|$ is some norm.

A subsequent definition removes details of initial conditions
and trjectories:  The sub-systems are synchronized with 
respect to the properties $\g_x$ and $\g_y$ if the previous 
definition holds on {\em all} trajectories.  The subsequent
definition is what many papers in the literature call
synchronization~\cite{chaos}.  However, as we and others have
shown, synchronization depends strongly on the 
trajectory~\cite{chaos,ashwin}.  Therefore, the trajectory 
dependence in the first definition can not be ignored.

We claim that this definition naturally follows from the intuitive
definition of synchronization, and that it encompasses all of the
interesting examples found in the literature.  A strength of this
definition is that the properties and comparison function are not
specified, {\em a priori}.  As shown below, different applications
{\em require} different properties and comparison functions, and 
those that are suitable for one application are often completely 
unsuitable for another.  For example, the following comparison
functions all appear in the literature
\begin{eqnarray}
\label{def_h}
\h[\g(\x), \g(\y)] & \equiv & \g(\x) - \g(\y) \\
\label{def_as_h}
\h[\g(\x), \g(\y)] & \equiv & \lim_{t \rightarrow \infty}
[\g(\x) - \g(\y)], \\
\label{def_av_h}
\h[\g(\x), \g(\y)] & \equiv & \lim_{T \rightarrow \infty} 
\frac{1}{T} \int_{t}^{t+T} [\g(\x(s)) - \g(\y(s))] ds .
\end{eqnarray}

Similar breadth occurs with properties.  The same two sub-systems
may be synchronized with respect to some properties, yet not 
synchronized with respect to other properties.  A lack of breadth 
is the Achilles heel of previous definitions.  Typically, they fail 
because they, {\em a priori}, specify properties and/or comparison
functions that must be applied to {\em all} types of synchronization.
Therefore, it is easy to find examples where the selected property
and/or comparison function is inappropriate.

Some may be concerned that the definition is too general.  We
argue, via analogy, that generality is also strength.  Perhaps 
the most useful concept in theoretical physics is a vector space.
The definition of a vector space is as general as the one proposed
for synchronization~\cite{bfmw}.  The definition does not specify
what constitutes a ``vector'' or the operation ``+''.
Thus, a range of things from matrices to Fourier series to bras
and kets are vectors in their respective vector spaces.  The 
definition only insists that the set of ``vectors'' obey a specific
series of abstract rules.  If a set obeys these rules then it is a
vector space, and the considerable power one obtains from that 
knowledge can be employed.  (Group theory is another example of an
extremely useful concept in physics whose definition is abstract.)
Our definition, gives an explicit list of four tasks and a condition
that must be satisfied for synchronization.  Like the definition of
a vector space (or a group) it provides a structural framework that
can be used for subsequent research.  We submit that this structure
is an improvement over the current situation of failed definitions
and enumerated lists. 

The remainder of this letter demonstrates the utility of the 
definition by discussing well-known examples.

\begin{center}
{\em Frequency Synchronization}
\end{center}
Sub-system properties used in frequency synchronization are
frequencies.  If the trajectory is mostly rotation about an 
axis then the measured frequencies ($\omega_x = \g(\x)$ and 
$\omega_y = \g(\y)$) are located at power spectra peaks
associated with the average rotation of the signal.  Examples
of such dynamics include, periodic motion, systems with phase
coherent chaotic attractors (like R\"{o}ssler~\cite{farmer}),
or systems with \v{S}ilnikov dynamics~\cite{jm}.  For these
examples, phase modulation contributes weakly to the dynamics,
and chaos (if it exists) results mainly from amplitude modulation.

The measurement function is typically $\h[\g(\x), \g(\y)] \equiv
n_x \omega_x - n_y \omega_y$ (where $n_x$ and $n_y$ are integers). 
Synchronization implies that the frequencies of the sub-systems
are commensurate
\begin{equation}
\label{freq}
n_x \omega_x - n_y \omega_y = 0.
\end{equation} 
Many text books discuss frequency synchronization for periodic 
systems~\cite{arnold}.

Frequency synchronization between coupled chaotic systems with
bi-stable attractors has also been examined~\cite{ask}. (The 
Lorenz and double scroll attractors are examples of bi-stable 
attractors.)  In Ref.~\cite{ask}, the properties are the average
frequency of switching between the two lobes of the attractors,
and frequency synchronization on a trajectory occurs if 
Eq.~(\ref{freq}) is satisfied.  Our definition works for this 
example.  Many other definitions fail because, either the
definition of phase is ambiguous~\cite{yl}, and/or $\| \x - \y 
\|$ need not remain small~\cite{cc}. 

Another example is frequency synchronization between a chaotic
bi-stable attractor and a periodic system.  The communication 
method of Hayes~{\em et. al.}~\cite{hgo} labels the lobes of
the attractor as 0 or 1, and uses small control signals to
produce a trajectory which encodes the message.  The obvious
choice for the sampling rate of the receiver is the mean
switching frequency of the chaotic system.  Therefore, a
periodic receiver with frequency $\omega_y$ is synchronized 
to a chaotic transmitter with switching frequency $\omega_x$
if $\omega_x = \omega_y$.  This type of synchronization fits 
into our definition, but does not seem to fit into any 
previous definition.

Frequency synchronization compares properties that are long
time averages of the trajectory. Therefore, it is a loose 
restriction on the dynamics of the sub-systems.  In particular,
it does not restrict the instantaneous values of the 
coordinates \x\ and \y.  All remaining examples compare 
properties whose measured values depend implicitly on time.

\begin{center}
{\em Phase Synchronization}
\end{center}
Phase synchronization involves sub-system properties called 
``phases''.  If the dynamics is chaotic and phase coherent then
one can introduce cylindrical coordinates, and unambiguously 
define the phase as the angle coordinate, $\phi(t)$.
However, other applications define the phase via a Hilbert
transform, in which case the phase may not be uniquely defined
on the sub-system~\cite{rpk}.  Also, there are examples where 
the measured phase, $\bfphi(t)$, is a vector 
obtained from a trajectory using none of the previous 
methods~\cite{yl}.

If the measured properties are given by $\g(\x) = \bfphi_{x}
(t)$ and $\g(\y) = \bfphi_y (t)$ then the most common 
comparison function is~\cite{rpk,rpk-1,gb},
\begin{equation}
\label{def_step}
\h[\g(\x),\g(\y)] = \mbf{U} \left[\mbf{\epsilon}, 
(\g(\x) - \g(\y)) \right].
\end{equation}
Here, $\mbf{U}(\mbf{u},\mbf{v})$ is a vector with $\alpha$-th
component $U_\alpha(\mbf{u},\mbf{v}) = \Theta \left[ u_\alpha
- |v_\alpha| \right]$, and $\Theta$ is the unit step function.
Equation~(\ref{def_step}) says that synchronization means
$|\phi_{x \alpha} - \phi_{y \alpha} | < \epsilon_\alpha$,
so $\mbf{\epsilon}$ is the maximum tolerable separation 
between the components of the phase.  The value of $\| 
\mbf{\epsilon} \|$ is usually small, but can not be set {\em
a priori} because its size is application dependent~\cite{gb}.
If ``phase slips'' occur then a comparison function using a 
time average like that of Eq.~(\ref{def_av_h}) is 
necessary~\cite{rpk}.

Phase synchronization 
only compares the phase variables.  In the synchronous state
the phases are locked, but the amplitudes can remain chaotic
and relatively uncorrelated.  Our definition includes phase
synchronization.  In contrast, definitions which focus on 
$\|\x - \y \|$~\cite{cc} fail because phase synchronization
does not restrict amplitudes.  Likewise, any definition that
forces one to use a specific type of phase will fail because
phase is not uniquely defined.

\begin{center}
{\em Identical Synchronization}
\end{center}
This is the most frequently discussed form of synchronization
within the nonlinear dynamics community~\cite{chaos}.  Here
the sub-systems are identical, and the properties are the
phase space variables, $\g(\x) = \x(t)$, and $\g(\y) = \y(t)$.
Most discussions in the literature use Eq.~(\ref{def_as_h}) as
the comparison function~\cite{chaos,cc}.

However, if the dynamics of the sub-systems are chaotic then
bursts (sudden loss and recovery of synchronous motion caused
by unstable periodic orbits within the attractor) occur on
chaotic trajectories for some forms of coupling~\cite{ashwin}.
Applications which can not tolerate bursts demand high quality
synchronization, where Eqs.~(\ref{def_step}) is used as the
comparison function~\cite{gb}.  If bursts are tolerable then 
a hybrid of Eqs.~(\ref{def_av_h}) and (\ref{def_step}) can be
used.

An engineering application called ``dead-beat'' synchronization
(only possible in discrete time dynamical systems) uses 
Eq.~(\ref{def_h}) as the comparison function~\cite{agt}.  This 
type of synchronization is often used to describe systems whose 
measured properties are restricted to a finite symbolic 
alphabet~\cite{neznam}. 

In their seminal paper, Afraimovich, Verichev, and 
Rabinovich~\cite{avr} generalized identical synchronization in
two different ways.  

\begin{center}
{\em Lag Synchronization}
\end{center}
Two sub-systems are lag synchronized if their measured properties
lag each other by a fixed amount of time, $\tau$.  A trivial
example is when the measured properties are $\g(\x) = \x(t)$
and $\g(\y) = \y(t+\tau)$, and the comparison function is 
Eq.~(\ref{def_h}).  For this example, the second sub-system 
follows the same trajectory as the first sub-system, but is 
$\tau$ units of time behind.

A nontrivial example is Ref.~\cite{rpk-1}.  In this paper the
measured properties are $\g(\x) = x_1(t)$ and $\g(\y) = y_1
(t+\tau)$ (the first components of \x\ and \y).  The comparison
function is $\h[\g(\x), \g(\y)] = K \left\langle [\g(\x) - \g
(\y)]^2 \right\rangle$, where $K$ is a constant and $\left\langle
\bullet \right\rangle$ is a time average.  For this example, the
sub-systems are {\em not} identical and $S^2(\tau) \equiv \|
\h \| = 0 $ for a {\em non-zero} value of $\tau$.  

Lag synchronization also occurs if, instead of a constant value of
$\tau$, one uses $\g(\y)=\y[T(t)]$, with $T: \R \to \R$ a 
homeomorphism with $\lim_{t \to \infty} \frac{T(t)}{t} = 
1$~\cite{avr}. 

The second generalization in Ref.~\cite{avr} follows from the
observation: If the sub-system are identical then the set $\x 
= \y$ defines an invariant manifold in the phase space of the 
large system.

\begin{center}
{\em Generalized Synchronization}
\end{center}
The literature is not consistent when discussing generalized
synchronization.  Most papers say that generalized
synchronization occurs if the measured properties are $\g(\x) =
\x$, $\g(\y) = \y$, and the comparison function, \h, is given by
\begin{equation}
\label{def_H}
\h[\g(\x), \g(\y)] = \H[\g(\x)] - \g(\y),
\end{equation}
where $\H$ is a smooth, invertible, time independent 
function~\cite{kp}.  Roughly speaking, the sub-systems are 
generally synchronized if $\y(t) = \H [\x(t)]$.  The equation,
$\y = \H(\x)$, defines an invariant manifold in the phase space
of the large system, and one can determine the state of one 
sub-systems from the state of the other sub-system~\cite{kp}.

However, Rulkov~{\em et. al.}~\cite{rsta} examined an example 
where the sub-systems have the same functional form but 
different parameter values.  Their numerical and experimental
evidence indicates that it is possible to have stable frequency 
synchronization on a trajectory and not have generalized 
synchronization in the sense discussed above.  For their example,
one sub-systems oscillated twice for every oscillation of the
other sub-system (i.e., $\omega_1/\omega_2 = 2$).  This implies
that it is impossible to construct a smooth invertible mapping 
$\y = \H(\x)$.  Therefore, the definitions in Ref.~\cite{kp}
fail.

This example illustrates that the definition of generalized 
synchronization needs to include \H's with a finite (or
perhaps countable) number of branches.   Similar conclusions
arise from Ref.~\cite{pyragas}.  Because we only require the
functional reationship, $\h[\g(\x), \g(\y)] = \zero$, between
the properties of the sub-system, both of these examples are
naturally contained within our definition of synchronization.
In particular, we do not insist that this relationship have
the form $\g(\y) = \H[\g(\x)]$.  Therefore, we argue that the
definition we have proposed is a more natural definition for
generalized synchronization.

In conclusion, we have describe four tasks that are required for
synchronization.  Based on this, we proposed a unified definition
of synchronization between finite dimensional dynamical systems.
We claim
that this definition encompasses all examples of synchronization
discussed in the literature, and that it offers a common language
and framework that can be used to discuss different types of
synchronization.

The authors would like the thank Drs. Lou Pecora, Tom Carroll, Ulli
Parlitz, Steve Strogatz, Brian Hunt, Dan Gauthier, and Jim Yorke for 
helpful discussions and suggestions.
This work was supported by an NSF CAREER grant number PHY-972236.


\begin{references}

\bibitem{chaos} {\em Chaos} {\bf 7} (1997); {\em IEEE Trans.
Circuits and Systems, part I} {\bf 44} (1997), and references
therein. 

\bibitem{pc} L. M. Pecora, and T. L. Carroll, {\em Phys. Rev.
Letts.} {\bf 64}, 821 (1990).

\bibitem{rpk} M. G. Rosenblum, A. S. Pikovsky, and J. Kurths,
{\em Phys. Rev. Letts. } {\bf 76}, 1804 (1996).

\bibitem{rpk-1} M. G. Rosenblum, A. S. Pikovsky, and J. Kurths,
{\em Phys. Rev. Letts. } {\bf 78}, 4193 (1997).

\bibitem{rsta} N. F. Rulkov, M. M. Sushchik, L. S. Tsimring, and 
H. D. I. Abarbanel, {\em Phys. Rev.} {\bf 51E}, 980 (1995);
N. F. Rulkov and M. M. Sushchik, {\em Phys. Lett.} {\bf 214A},
145 (1996).

\bibitem{blek}  I. I. Blekhman, P.S. Landa, and M. G. Rosenblum, 
{\em Appl. Mech. Rev.} {\bf 48}, 733 (1995). 

\bibitem{arnold} See the discussion of Arnold tongues in 
E. Ott, {\em Chaos in dynamical systems}
(Cambridge University Press, New York, NY, 1994).

\bibitem{tdhh} D. Y. Tang, R. Dykstra, M. W. Hamilton, and
H. R. Heckenberg, {\em Chaos} {\bf 8}, 697 (1998).

\bibitem{josic} K. Josic, {\em Phys. Rev. Letts.} {\bf 80}, 
3054 (1998).

\bibitem{cc} P. F. Curran and L. O. Chua, {\em Int. J. Bif. and
Chaos} {\bf 7}, 1375 (1997); and papers in 
Ref.~\protect{\cite{chaos}}.

\bibitem{kp} L. Kocarev and U. Parlitz, {\em Phys. Rev. Letts.}
{\bf 76}, 1816 (1996); U. Parlitz, L. Junge, and L. Kocarev, 
{\em Phys. Rev. Letts.} {\bf 79}, 3158 (1997); B. R. Hunt, E. 
Ott, and J. A. Yorke, {\em Phys. Rev.} {\bf 55E}, 4029 (1997);
R. Brown, {\em Phys. Rev. Letts.} {\bf 81}, 4835 (1998).

\bibitem{oxford} A. S. Hornby, {\em Oxford Advanced Dictionary}, 
(Oxford University Press, 1974).  

\bibitem{ashwin} P. Ashwin, J. Buescu, and I. Stewart {\em Phys.
Lett.} {\bf 193A}, 126 (1994); P. Ashwin, J. Buescu, and I. 
Stewart  {\em Nonlinearity} {\bf 9}, 703 (1996); R. Brown and 
N. F. Rulkov, {\em Phys. Rev. Letts.} {\bf 78}, 4189 (1997); 
R. Brown and N. Rulkov, {\em Chaos} {\bf 7}, 395 (1997).  

\bibitem{bfmw} F. W. Byron and R. W. Fuller, {\em Mathematics of
Classical and Quantum Physics}, (Addison-Wesley, 1969), p~88;  
J. Mathews and R. L. Walker, {\em Mathematical Methods of Physics},
(W. A. Benjamin, Inc., 1964), p~140.

\bibitem{farmer} J. D. Farmer, {\em Phys. Rev. Letts. } {\bf 47},
179 (1981); J. Crutchfield, D. Farmer, N. Packard, R. Shaw, G. Jones, 
and R. J. Donnelly, {\em Phys. Lett.} {\bf 76A}, 1 (1980); E. F.
Stone, {\em Phys. Lett.} {\bf 163A}, 367 (1992) .

\bibitem{jm} K. Judd and A. Mees, {\em Physica} {\bf 82D}, 426 (1995).

\bibitem{ask} V. S. Anishchenko, A. N. Silchenko, and I. A. Khovanov,
{\em Phys. Rev.} {\bf 57E}, 316 (1998).

\bibitem{yl} T. Yalcinkaya and Y-C Lai, {\em Phys. Rev. Letts.} {\bf
79}, 3885 (1997).

\bibitem{hgo} S. Hayes, C. Grebogi, and E. Ott, {\em Phys. Rev. Letts}
{\bf 70}, 3031 (1993).

\bibitem{gb} D. J. Gauthier, and J. C. Bienfang, {\em Phys. Rev. 
Letts.} {\bf 77}, 1751 (1996).

\bibitem{agt} A. De Angeli, R. Genesio, and A. Tesi, {\em IEEE Trans.
Circuits and Systems, part I} {\bf  42}, 54 (1995). 

\bibitem{neznam} W. E. Savage, {\em Bell Syst. Tech. J.} {\bf 46}, 449 

\bibitem{avr} V. S. Afraimovich, N. N. Verichev, and M. I. Rabinovich, {\em
Radiophys. Quantum Elect.} {\bf 29}, 747 (1986).

\bibitem{pyragas} K. Pyragas, {\em Phys. Rev.} {\bf 54E}, R4508 (1996).

\end{references}
\end{document}